\documentclass[manuscript]{aastex}

\def \ms{ms$^{-1}$}
\def \msun{$M_{\odot}$}
\def \rsun{$R_{\odot}$}
\def \lsun{$L_{\odot}$}

\def \mearth{$M_{\oplus}~$}
\def \me{$M_{\oplus}$}
\def \re{$R_{\oplus}$}

\slugcomment{}

\shorttitle{A Super-Earth in the HZ of GJ~581}
\shortauthors{Vogt et al.}

\begin{document}

\title{The Lick-Carnegie Exoplanet Survey: A 3.1\,\mearth Planet in the Habitable Zone of the Nearby M3V Star Gliese 581}

\author{Steven S. Vogt\altaffilmark{1}, R. Paul Butler\altaffilmark{2}, E. J. Rivera\altaffilmark{1}, N. Haghighipour\altaffilmark{3}, Gregory W. Henry\altaffilmark{4}, and Michael H. Williamson\altaffilmark{4}}

\altaffiltext{1}{UCO/Lick Observatory, University of California, Santa Cruz, CA 95064}
\altaffiltext{2}{Department of Terrestrial Magnetism, Carnegie Institution of Washington, 5241 Broad Branch Road, NW, Washington, DC 20015-1305}
\altaffiltext{3}{Institute for Astronomy and NASA Astrobiology Institute, University of Hawaii-Manoa, Honolulu, HI 96822}
\altaffiltext{4}{Tennessee State University, Center of Excellence in Information Systems, 3500 John A. Merritt Blvd., Box 9501, Nashville, TN. 37209-1561}

\begin{abstract}
We present 11 years of HIRES precision radial velocities (RV) of the nearby M3V
star Gliese 581, combining our data set of 122 precision RVs with
an existing published 4.3-year set of 119 HARPS precision RVs. The
velocity set now indicates 6 companions in Keplerian motion around
this star. Differential photometry indicates a likely stellar rotation period of $\sim94$ days
and reveals no significant periodic variability at any of the Keplerian periods, supporting
planetary orbital motion as the cause of all the radial velocity variations. The combined data set strongly
confirms the 5.37-day, 12.9-day, 3.15-day, and 67-day planets previously
announced by \cite{bon05}, \cite{udr07}, and \cite{may09}. The observations also
indicate a 5th planet in the system, GJ~581f, a minimum-mass 7.0\,\mearth planet
orbiting in a 0.758 AU orbit of period 433 days and a 6th planet, GJ~581g, a
minimum-mass 3.1\,\mearth planet orbiting at 0.146 AU with a period of 36.6
days. The estimated equilibrium temperature of GJ~581g is 228~K, placing it
squarely in the middle of the habitable zone of the star and offering
a very compelling case for a potentially habitable planet around a
very nearby star. That a system harboring a potentially habitable planet
has been found this nearby, and this soon in the relatively early history of
precision RV surveys, indicates that $\eta_{\oplus}$, the fraction
of stars with potentially habitable planets, is likely to be substantial. This
detection, coupled with statistics of the incompleteness of present-day
precision RV surveys for volume-limited samples of stars in the immediate solar
neighborhood suggests that $\eta_{\oplus}$ could well be on the order of a few
tens of percent.  If the local stellar neighborhood is a representative sample of
the galaxy as a whole, our Milky Way could be teeming with potentially habitable
planets.

\end{abstract}

\keywords{stars: individual: GJ~581 HIP 74995 -- stars: planetary systems -- astrobiology}

\section{Introduction}

There are now nearly 500 known extrasolar planets, and discovery work continues
apace on many fronts: by radial velocities (RV), gravitational microlensing,
transit surveys, coronography, nulling interferometry, and astrometry. By far
the most productive discovery technique to date has been through the use of
precision RVs to sense the barycentric reflex velocity of the host star
induced by unseen orbiting planets. In recent years, the world's leading RV
groups have improved precision down to the $\sim$1 \ms\ level, and even below,
extending detection levels into the range of planets with masses less than
10\,\me, commonly referred to as ``Super-Earths''. This level of precision is
now bringing within reach one of the holy grails of exoplanet research, the
detection of $\sim$Earth-size planets orbiting in the habitable zones (HZ) of
stars. Nearby K and M dwarfs offer the best possibility of such detections, as
their HZ's are closer in, with HZ orbital periods in the range of weeks to
months rather than years. These low mass stars also undergo larger reflex
velocities for a given planet mass. To this end, we have had a target list of
$\sim$400 nearby quiet K and M dwarfs under precision RV survey with HIRES at
Keck for the past decade.

One of these targets, the nearby M3V star GJ~581 (HIP 74995), has received
considerable attention in recent years following the announcement by
\cite{bon05}, hereafter Bonfils05, of a 5.37-day hot-Neptune (GJ~581b, or
simply planet-b) around this star. More recently, the Geneva group
\citep{udr07}, hereafter Udry07, announced the detection of two additional
planets (c and -d) in this system, one close to the inner edge of
the HZ of this star and the other close to the outer edge. Planet-c was
reported to have a period of 12.931 days and $m\sin{i}= 5.06$\,\mearth
whereas planet-d was reported to have a period of 83.4 days and
$m\sin{i}= 8.3$\,\me.

The Geneva group's announcement of planet-c generated considerable excitement
because of its small minimum mass (5\,\me, well below the masses of the ice
giants of our solar system and potentially in the regime of rocky planets or
Super-Earths) and its location near the inner edge of the HZ of this star. An
assumed Bond albedo of 0.5 yielded a simple estimate of $\sim$320~K for the
equilibrium temperature of the planet, suggesting the possibility that it was a
habitable Super-Earth. However, a more detailed analysis by \cite{sel07}, that
included the greenhouse effect and the spectral energy distribution of GJ~581,
concluded that planet-c's surface temperature is much higher than the
equilibrium temperature calculated by Udry07 and that it is unlikely to host
liquid water on its surface. \cite{sel07} concluded that both planets c and d
are demonstrably outside the conservative HZ of this star, but that given a
large atmosphere, planet-d could harbor surface liquid water. \cite{chy07}
reached a similar conclusion that neither planets c nor d is in the HZ, but
that planet-d could achieve habitability provided a greenhouse effect of 100~K
developed. Moreover, if these planets are tidally spin-synchronized, planet-c
could conceivably have atmospheric circulation patterns that might support
conditions of habitability. \cite{von07} also concluded that planet-c is too
close to the star for habitability. They argue, however, that if planet-d has
a thick atmosphere and is tidally locked, it may lie just within the outer edge
of the HZ. Both \cite{von07} and \cite{sel07} conclude that planet-d would be
an interesting target for the planned TPF/Darwin missions.

\cite{beu08} studied the dynamical stability and evolution of the GJ~581 system
using the orbital elements of Udry07, which they integrated forward for
$10^{8}$ years. They observed bounded chaos (see e.g. \cite{las97}), with
small-amplitude eccentricity variations and stable semi-major axes. Their
conclusions were unaffected by the presence of any as-yet-undetected outer
planets. On dynamical stability grounds, they were able to exclude inclinations
$i \leq 10^{\circ}$ (where $i=0^{\circ}$ is face-on). 

Last year, \cite{may09}, hereafter Mayor09, published a velocity update wherein
they revised their previous claim of an 8 \mearth planet orbiting with an
83-day period, to a 7.1 \mearth planet orbiting at 67-days, citing confusion
with aliasing for the former incorrect period. Mayor09 also reported another
planet in the system at 3.148 days with a minimum mass of 1.9 \me. They
also presented a dynamical stability analysis of the system. In particular, the
addition of the 3.15d planet, GJ~581e, greatly strengthened the inclination
limit for the system.  The planet was quickly ejected for system inclinations
less than 40$^{\circ}$. This dynamical stability constraint implies an upper
limit of 1.6 to the $1/\sin{i}$ correction factor for any planet's minimum mass
(assuming coplanar orbits). Most recently, \cite{daw10} published a detailed
study of the effects of aliasing on the GJ~581 data set of Mayor09. They
concluded that the 67-day period of GJ~581c remains ambiguous, and favored a
period of 1.0125 days that produced aliases at both 67 days and 83 days.

The Gliese 581 system exerts an outsize fascination when compared to many of the
other exoplanetary systems that have been discovered to date. The interest stems
from the fact that two of its planets lie tantalizingly close to the expected
threshold for stable, habitable environments, one near the cool edge, and one
near the hot edge. We have had GJ~581 under survey at Keck Observatory for over
a decade now. In this paper, we bring 11 years of HIRES precision RV data to
bear on this nearby exoplanet system. Our new data set of 122 velocities, when
combined with the previously published 119 HARPS velocities, effectively doubles
the amount of RVs available for this star, and almost triples the time base of
those velocities from 4.3 years to 11 years. We analyze the combined precision
RV data set and discuss the remarkable planetary system that they reveal.

\section{Radial Velocity Observations}

The RVs presented herein were obtained with the HIRES spectrometer
\citep{vog94} of the Keck I telescope. Typical exposure times on GJ~581 were
600 seconds, yielding a typical S/N ratio per pixel of 140. Doppler shifts are measured by placing
an Iodine absorption cell just ahead of the spectrometer slit in the converging
f/15 beam from the telescope. This gaseous absorption cell superimposes a rich
forest of Iodine lines on the stellar spectrum, providing a wavelength
calibration and proxy for the point spread function (PSF) of the spectrometer.
The Iodine cell is sealed and temperature-controlled to 50 $\pm$ 0.1 C such
that the column density of Iodine remains constant \citep{but96}. For the Keck
planet search program, we operate the HIRES spectrometer at a spectral
resolving power R $\approx$ 70,000 and wavelength range of 3700\,--\,8000\,\AA,
though only the region 5000\,--\,6200\,\AA\ (with Iodine lines) is used in the
present Doppler analysis. Doppler shifts from the spectra are determined with
the spectral synthesis technique described by \cite{but96}. The Iodine region
is divided into $\sim$700 chunks of 2\,\AA\ each. Each chunk produces an
independent measure of the wavelength, PSF, and Doppler shift. The final
measured velocity is the weighted mean of the velocities of the individual
chunks.

In August 2004, we upgraded the focal plane of HIRES to a 3-chip CCD mosaic of
flatter and more modern MIT-Lincoln Labs CCD's. No zero point shift in our RV
pipeline was incurred from the detector upgrade. Rather, the new CCD mosaic
eliminated a host of photometric problems with the previous Tek2048 CCD
(non-flat focal plane, non-linearity of CTE, charge diffusion in the silicon
substrate, overly-large pixels, and others). The deleterious effects of all
these shortcomings can be readily seen as larger uncertainties on the
pre-August 2004 velocities.

In early 2009, we submitted a paper containing our RVs up to that date for GJ~581 that
disputed the 83-day planet claim of Mayor09. One of the referees (from the
HARPS team) kindly raised the concern (based partly on our larger value for
apparent stellar jitter) that we may have some residual systematics
that could be affecting the reliability of some of our conclusions.
In the precision RV field there are no suitable standards by which teams can
evaluate their performance and noise levels; so, it is rare but also extremely
useful for teams to be able to check each other using overlapping target stars,
like GJ~581, for inter-comparison. So, we took the HARPS team's concerns to
heart and withdrew our paper to gather another season of data, to do a
detailed reanalysis of our uncertainty estimates, and to scrutinize our 15-year
1500-star data base for evidence of undiscovered systematic errors.

Soon after we withdrew our 2009 paper, Mayor09 published a revised model
wherein they altered their 83-day planet period to 66.8 days (citing confusion
by yearly aliases) and also announced an additional planet in the system near
3.15 days. For our part, as a result of our previous year's introspection, we
discovered that the process by which we derive our stellar template spectra was
introducing a small component of additional uncertainty that added
about 17\% to our mean internal uncertainties. This additional noise source
stems from the deconvolution process involved in deriving stellar template
spectra. This process works quite well for G and K stars, but it is prone to
extra noise when applied to heavily line-blanketed M dwarf spectra. We have
included this in our present reported uncertainties for GJ~581, and are working on
improvements to the template deconvolution process. Furthermore, our
existing template for this star, taken many years ago, was not up to the task
of modeling RV variation amplitudes down in the few \ms\ regime. So, over the
past year, we obtained a much higher quality template for GJ~581. 

The HIRES velocities of GJ~581 are presented in Table 1, corrected to the solar
system barycenter. Table 1 lists the JD of observation center, the RV, and the internal
uncertainty. The reported uncertainties reflect only one term in the overall
error budget, and result from a host of systematic errors from characterizing
and determining the PSF,  detector imperfections, optical aberrations, effects
of under-sampling the Iodine lines, etc. Two additional major sources of error
are photon statistics and stellar jitter. The former is already included in our
Table 1 uncertainties. The latter varies widely from star to
star, and can be mitigated to some degree by selecting magnetically-inactive
older stars and by time-averaging over the star's unresolved low-degree surface
p-modes. The best measure of overall precision for any given star is simply to
monitor an ensemble of planet-free stars of similar spectral type,
chromospheric activity, and apparent magnitude, observed at similar cadence and
over a similar time base.  Figures 2, 3, and 4 of Butler et al. (2008) show 12
M dwarfs with B-V, V magnitude, and chromospheric activity similar to GJ~581.
In any such ensemble, it is difficult to know how much of the root-mean-square
(RMS) of the RVs is due to as-yet-undiscovered planets and to stellar jitter.
However, these stars do establish that our decade-long precision is better than
3 \ms\ for M dwarfs brighter than V=11, including contributions from stellar
jitter, photon statistics, undiscovered planets, and systematic errors.

\section{Properties of GJ~581}

The basic properties of GJ~581 were presented by Bonfils05 and Udry07 and will,
for the most part, simply be adopted here. Briefly recapping from Bonfils05 and
Udry07, GJ~581 is an M3V dwarf with a parallax of 159.52 $\pm$ 2.27 mas
(distance of 6.27 pc) with V = 10.55 $\pm$ 0.01 and B-V = 1.60. The parallax
and photometry yield absolute magnitudes of M$_{\rm V}$ = 11.56 $\pm$ 0.03 and
M$_{\rm K}$ = 6.86 $\pm$ 0.04. The V-band bolometric correction of 2.08
\citep{del98} yields a luminosity of 0.013 \lsun. The K-band mass-luminosity
relation of \cite{del20} indicates a mass of 0.31 $\pm$ 0.02 \msun, and the
mass-radius relations of \cite{cha20} yield a radius of 0.29 \rsun.
\cite{bea06} report the [Fe/H] of GJ~581 to be -0.33, while Bonfils05 report
[Fe/H] = -0.25. Both results are consistent with the star being slightly
metal-poor, in marked contrast to most planet-bearing stars that are of
super-solar metallicity. \cite{joh09} presented a broadband
(V-K) photometric metallicity calibration for M dwarfs that, in conjunction
with the star's broadband magnitudes implies a metallicity of [Fe/H] = -0.049.
Most recently, \cite{rojas10} estimated the metallicity at -0.02, while \cite{sch10}
cite a metallicity of -0.22. Thus, GJ~581 appears to be basically of solar or slightly sub-solar
metallicity, yet has produced at least 4 or more low-mass planets. However,
this is no cause for surprise. \cite{lau04}  and \cite{ida05} have argued that
the formation of low-mass planets should not be unduly affected by modestly
subsolar metallicity.

Udry07 report GJ~581 to be one of the least active stars on the HARPS M-dwarf
survey, with Bonfils05 reporting line bisector shapes stable down to their
measurement precision levels. Udry07 report a measured
$v\sin{i} \leq 1$ kms$^{-1}$. They thus find GJ~581 to be quite inactive with
an age of at least 2 Gyr. Our measurement of $\log{R'_{hk}} = -5.39$ leads to
an estimate \citep{jtw05} of 1.9 \ms\ for the expected RV jitter due to
stellar surface activity and an age estimate of 4.3 Gyr. 

\section{Photometric Observations}

Precise photometric observations of planetary host candidate stars are 
useful to look for short-term, low-amplitude brightness variability due 
to rotational modulation in the visibility of starspots and plages
\citep[see, e.g.,][]{hfh95}.  Long-term brightness monitoring of 
these stars enabled by our automatic telescopes can detect brightness
changes due to the growth and decay of individual active regions as well
as brightness variations associated with stellar magnetic cycles 
\citep{h99,lsh07,hhl09}.  Therefore, photometric observations of planetary 
candidate stars help to determine whether the observed radial velocity 
variations are caused by stellar activity (spots and plages) or reflex 
motion due to the presence of orbiting companions.  \citet{qetal01} and 
\citet{psc04} have documented several examples of solar-type stars whose 
periodic radial velocity variations were caused by stellar activity. 

GJ~581 has also been classified as the variable star HO Librae, though
\cite{wei94} reported its short-term variability to be at most 0.006
magnitudes. Udry07 report the star to be constant to within the 5 millimag
Geneva photometry catalog precision of V=10.5 stars.

We acquired new photometric observations of GJ~581 in the Johnson V band 
during the 2007 and 2008 observing seasons with an automated 0.36~m 
Schmidt-Cassegrain telescope coupled to an SBIG ST-1001E CCD camera.  
This Tennessee State University telescope was mounted on the roof of 
Vanderbilt University's Dyer Observatory in Nashville, Tennessee.

Differential magnitudes were computed from each CCD image as the difference
in brightness between GJ~581 and the mean of four constant comparison stars
in the same field. A mean differential magnitude was computed from
usually ten consecutive CCD frames. Outliers from each group of ten
images were removed based on a $3\sigma$ test.  If three or more outliers
were filtered from any group of ten frames (usually the result of
non-photometric conditions), the entire group was discarded. One or two 
mean differential magnitudes were acquired each clear night; our final data 
set consists of 203 mean differential magnitudes spanning 530 nights.

Our 203 photometric observations are plotted in the top panel of Figure~1;
they scatter about their mean with a standard deviation of 0.0049 mag.
A periodogram of the observations, based on least-squares sine fits, is 
shown in the second panel, resulting in a best-fit period of $94.2~\pm~1.0$ 
days. That rotation period is quite similar to the rotational period of another
important M dwarf planet host, GJ 876, and gives added confidence to the
current findings. It is also consistent with GJ 581's  low activity and age
estimate. In the third panel, we plot the observations modulo the 94.2-day 
photometric period, which we take to be the star's rotation period.  A 
least-squares sine fit on the rotation period gives a semi-amplitude of 
$0.0030~\pm~0.0004$ mag. The window function for the rotation period is
plotted in the bottom panel. Five of the six radial velocity periods discussed
below are indicted by vertical dotted lines in the second and fourth
panels; our data set is not long enough to address the 433-day period
of GJ~581f. As will be shown below, none of the five periods coincide
with any significant dip in the periodogram.

\begin{figure}
\epsscale{.70}
\plotone{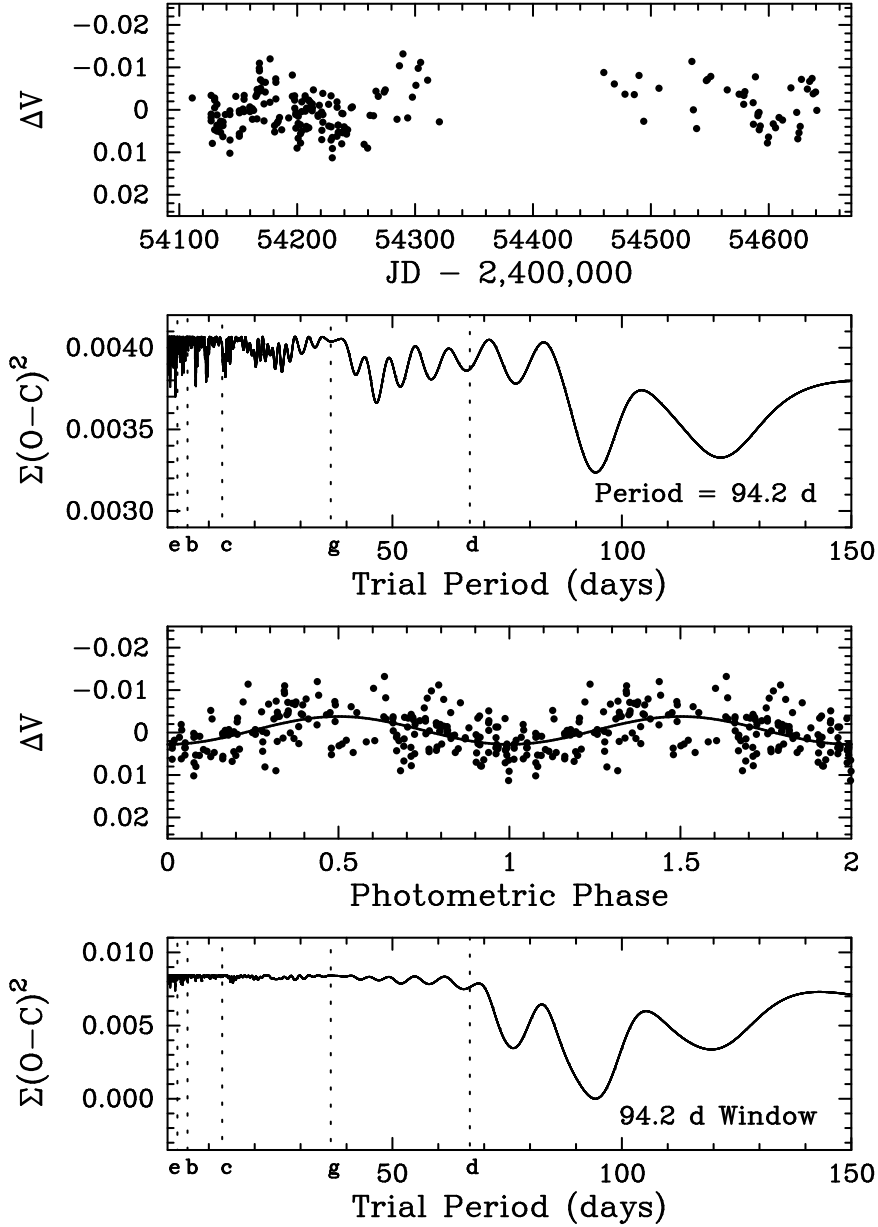}
\caption{($Top$):  Photometric $V$-band observations of GJ~581 acquired 
during the 2007 and 2008 observing seasons with an automated 0.36~m imaging 
telescope.  ($Second~Panel$):  Periodogram analysis of the observations gives 
the star's rotation period of 94.2 days. ($Third~Panel$):  The photometric 
observations phased with the 94.2-day period reveal the effect of rotational 
modulation in the visibility of photospheric starspots on the brightness of 
GJ~581. ($Bottom$): Window function of the 94.2-day rotation period.  
The radial velocity periods of 5 of the 6 planetary companions are indicated
by vertical dotted lines in the second and fourth panels.}
\end{figure}

\section{Orbital Analysis}

We obtained 122 RVs with the HIRES spectrometer at Keck. The data set spans
10.95 years with a peak-to-peak amplitude of 37.62 \ms, an RMS velocity
scatter of 9.41 \ms, and a mean internal uncertainty of 1.70 \ms. Figure 2
(top panel) presents the RVs tabulated in Table 1, combined with the HARPS
RVs published by Mayor09. The 122 (red) hexagon points are the HIRES
observations, while the HARPS observations are shown as (blue) triangle
points. A zero-point offset of 1.31 \ms\ was removed between the two data
sets, and Figure 2 has this offset included. The HARPS data consist of
119 observations at a reported median uncertainty of 1.10 \ms\ and
extending over 4.3 years. The peak-to-peak amplitude of the HARPS data
set is 39.96 \ms. The combined data set has 241 velocities, with a median
uncertainty of 1.30 \ms.

\begin{figure}
\includegraphics[angle=-90,scale=0.55]{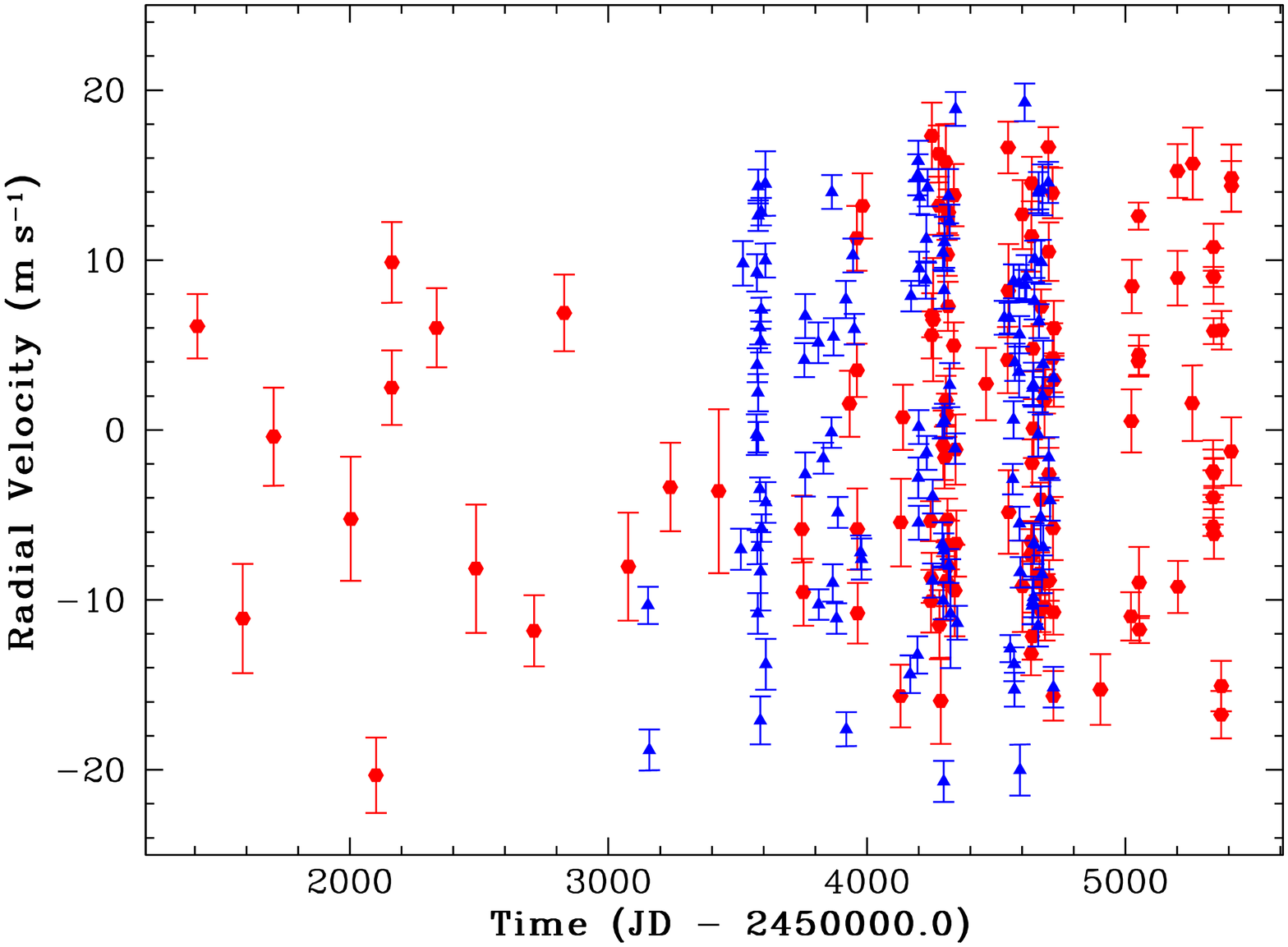}
\includegraphics[angle=-90,scale=0.55]{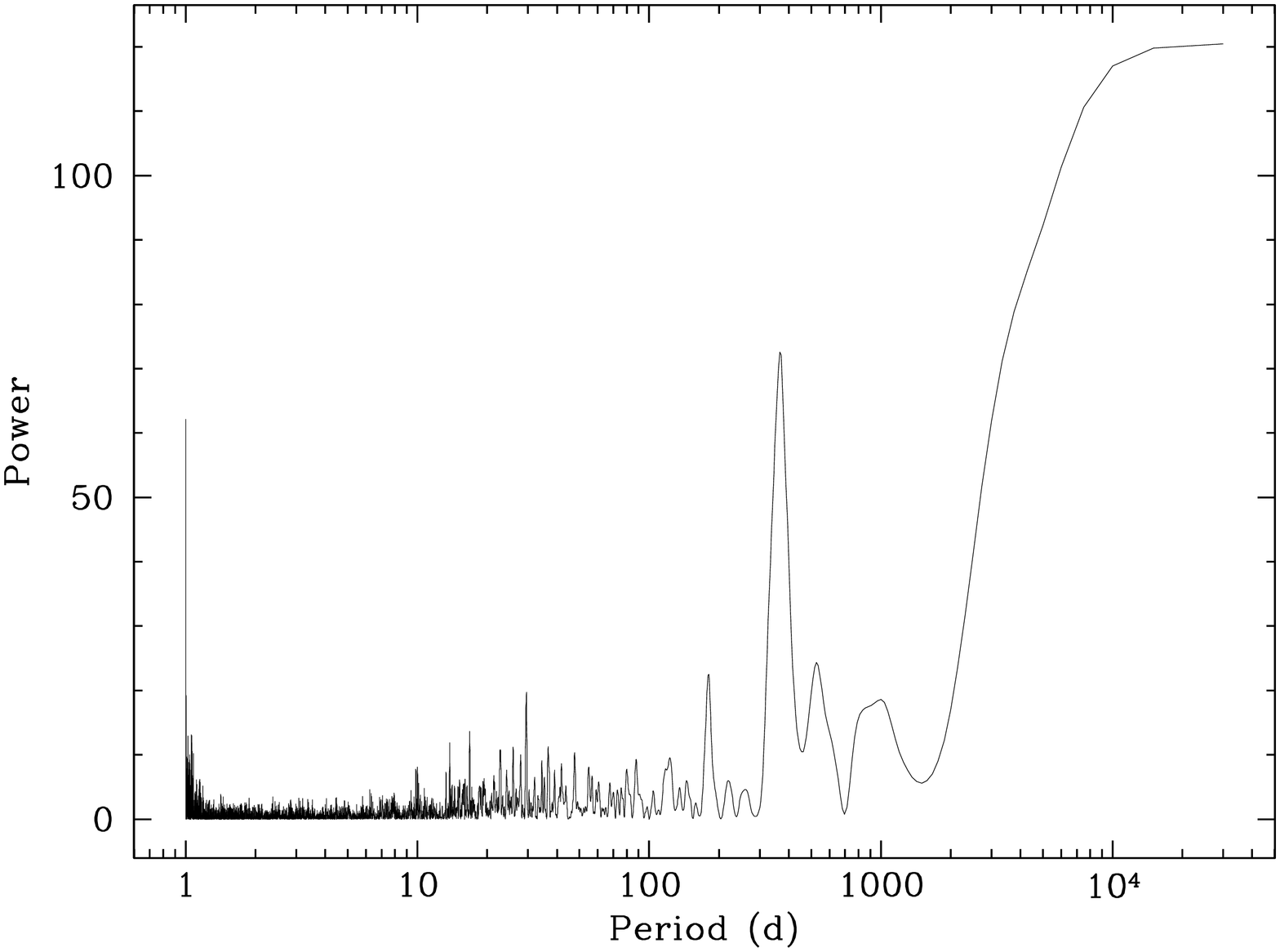}
\caption{Top panel: Combined RV data of GJ~581 from HIRES (red hexagons) and HARPS (blue triangles). Lower panel: spectral window }
\end{figure}

For the orbital fits, we used the SYSTEMIC Console \citep{mes09,mes10}. We
assume coplanar orbits with $i=90^{\circ}$ and
$\Omega=0^{\circ}.$ Uncertainties are based on 1000 bootstrap trials. We take
the standard deviations of the fitted parameters to the bootstrapped RVs as
the uncertainties in the fitted parameters. The fitted mean anomalies are
reported at epoch JD 2451409.762. The assumed mass of the central star is
0.31$M_{\odot}.$ For all fits presented here, we fixed the eccentricities
at zero since the amplitudes are all quite small and extensive modeling
revealed that allowing eccentricities to float for any or all of the 6 planets does
not significantly improve the overall fit.

The power spectrum of the sampling window is shown in the lower panel of
Figure 2. As expected, there is some spurious power created by the
sampling times near periods of 1.003d (the solar day in sidereal day units), 29.5d (the lunar
synodic month), 180d ($\sim$1/2 year), and 364d ($\sim$1 year), all
artifacts of the nightly, monthly, and yearly periods on telescope scheduling.

\begin{figure}
\epsscale{0.95}
\plotone{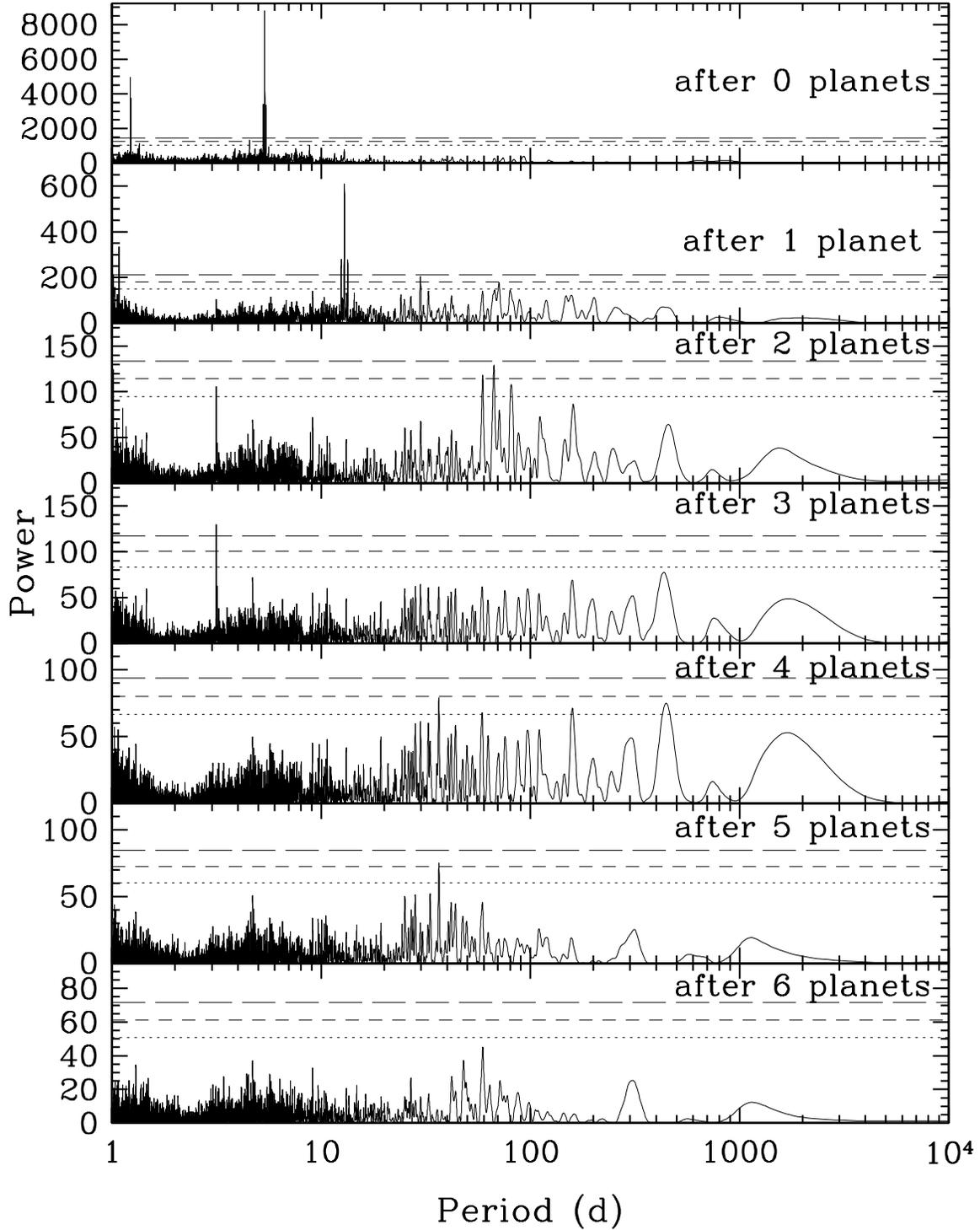}
\caption{From top to bottom, power spectra of the residuals to the 0-, 1-, 2-, 3-, 4-, 5-, and 6-planet solutions, respectively. The horizontal lines in each periodogram roughly indicate the 0.1\%, 1.0\%, and 10.0\% False Alarm Probability (FAP) levels from top to bottom.}
\end{figure}

The top panel of Figure 3 shows the power spectrum of the RV data.
Following \cite{gil87} (hereafter GB87), in Figure 3, we use an
error-weighted version of the Lomb-Scargle periodogram.
The horizontal lines in the periodograms in Figure 3 roughly
indicate the 0.1\%, 1.0\%, and 10.0\% False Alarm Probability (FAP)
levels from top to bottom. To determine better estimates of the FAPs
of the prominent peaks in the periodograms, we define the
noise-weighted power in a prominent peak with (GB87) 
\begin{equation}
p_{0} = {N\over{4}}{x_0^2 \over{\sigma_0^2}},
\end{equation}
\noindent
where N is the number of observations, $x_0$ is the RV half-amplitude
implied by the peak, and $\sigma_0^2$ is the variance in the data or
residuals prior to fitting out the implied planet. Additionally, we
can also define power in a prominent peak as (\citet{Cum04}):
\begin{equation}
p_0 = {(N-2)\over{2}} {(\chi_{\rm constant}^2 - \chi_{\rm circ}^2) \over {\chi_{\rm circ}^2}},
\end{equation}
\noindent
where $\chi_{\rm circ}^2$ is the reduced chi-squared for a circular
fit at/near the period implied by the peak and $\chi_{\rm constant}^2$
is the reduced chi-squared for a constant RV model of the data or residuals.

Estimation of the false-alarm probability of a given peak requires
knowledge of the number of independent frequencies, $M$ in the data
set. Given the highly uneven sampling, $M$ considerably exceeds our
$N=241$ Doppler velocity measurements. Using the Monte-Carlo procedure
outlined by \cite{pre92}, we find that $M=2525$. 

The FAP is the chance that a peak as high as, or higher than, that
observed in the periodogram would occur by chance, 
\begin{equation}
Pr(p_0,M) = 1-[1-\exp(-p_0)]^M\,.
\end{equation}
\noindent
In general, we find that $M$ is roughly the same for both definitions of
$p_0$ above.

Note that there are discrepancies between our FAPs quoted below
and the FAP lines shown in Figure 3. Here we explain the reasons
for these discrepancies. The (raw) power levels shown in Figure 3
are based on Equations 1 and 2 in GB87. The FAP lines are based
on the method to calculate the number of degrees of freedom, 
$M$, suggested in Section 13.7 of \cite{pre92}, except that we
assume a Gaussian distribution with a standard deviation equal to
the velocity scatter of the data or residuals. However, the FAPs we quote below
for each fitted planet are for power levels defined by Equation 2 above.

Figure 3 shows the power spectra of the residuals of the RV data from
the best Keplerian fits for models with $n$ planets (with $n$ ranging
from 0 to 6). The eccentricities are held fixed at 0 throughout the
fitting process. The dominant spike in the top panel is at 5.368 days
and is the well-known Hot-Neptune (GJ~581b) first reported by Bonfils05.
The power implies a minimum-mass $m\sin{i}=$15.6\,\mearth companion in
a 0.041 AU orbit. The reduced chi-squared statistic (using 5 free parameters) for this 1-planet
fit is 8.426, with an RMS of 3.65 \ms. The estimated FAP is
$6.8\times10^{-306},$ in keeping with the extremely strong detection.

The second panel down in Figure 3 shows the power spectrum of the residuals to
the 1-planet fit. This power spectrum is dominated by a peak at 12.92 days.
A 2-planet fit for the 12.92-day peak (planet-c first reported by Udry07)
reveals a minimum-mass 5.5 \mearth planet in a 0.073 AU orbit. The 2-planet fit
achieves a reduced chi-squared statistic (using 8 free parameters) of 4.931, and an RMS of 2.90 \ms. The
estimated FAP is $2.3\times10^{-33}$.  So, the 12.92-day planet-c first reported
by Udry07 also seems well-confirmed.

The third panel down of Figure 3 shows the power spectrum of the residuals
of the 2-planet model. As Mayor09 found, the next obvious peak to fit is the
maximum peak in the group near 67 days. Mayor09 found that this group is a set
of 3, with the true peak at 67 days, and 1-year aliases near 59 and 82 days
$(1/67 - 1/365 \sim 1/82$, and $1/67 + 1/365 \sim 1/57$). We explored various fitting
branches involving the 59d and 82d peaks for planet d. Fitting
for the 59-day peak left pronounced residuals at both 67 and 82 days. Fitting out the
82-day peak left pronounced residual peaks near 59 days, 37 days and 158 days. Neither the
59-day nor the 82-day fitting branches led to final solutions that were as good as the
67-day branch. We therefore concur with Mayor09 that the 67-day is the correct choice for
planet d. A fit to the 66.9-day peak indicates a minimum-mass 4.4\,\mearth planet in a
0.218 AU orbit. The 3-planet fit results in a reduced chi-squared statistic (using 11 free parameters) of
4.207, with an RMS of 2.72 \ms. The estimated FAP is  $2.5\times10^{-6}$. Thus,
the 67-day 3rd planet announced by Mayor09 seems well-supported by the present
data set.

At this point, there are also similar-power peaks present very near 1.00 day,
both above and below. These ``near-1-day'' peaks appear frequently in our RV
data sets and typically arise from aliasing effects, as discussed in detail by
\cite{daw10}. They are due partly to the fact that exoplanet observations are done only
at night. \cite{daw10} looked carefully at the HARPS data set for GJ 581 and concluded
that it remains unclear whether the period of GJ 581d is 67 days, or 83 days,
or even their preferred value of 1.0125 days, and that further observations were required to resolve
the ambiguity. In our experience, RV power from a star being orbited by legitimate
planets roughly in the 20\,--\,90 period range can feed substantial amounts of
that power into peaks very near 1.00 day, by beating with the sidereal and
solar days. Thus, while it may be possible on rare occasion to encounter a true
planet orbiting a given star with a period very near 1.00 day, this will be the
exceptional case, and not very compelling from a purely Bayesian point of view.
In addition, one can only use this alternative once in a system to explain away
a suspected planet peak up at a longer period. Multiple longer period peaks
would require multiple planets at or very near 1.00 day, and that is dynamically untenable.

To look into this more carefully, we intentionally obtained some extended
cadence over the course of nights on May 21-25, June 21-23, and
again on July 30-31, 2010. We then carefully examined the periodogram
of the residuals of the two-planet fit. The periodogram has two prominent
peaks at 66.9645 days and 1.0126 days with raw powers of 129.070
and 124.310, respectively. The ratio of the power levels is 1.038.
We generated mock RV sets based on two models. First, we took the
three-planet fit with the third planet at 1.0126 days and scrambled
the residuals 1000 times. We fit two planets to each mock RV set.
We then examined the periodograms of the residuals. In particular, we
measured how frequently the ratio of the power levels at the two periods
exceeds 1.038. Then we repeated this procedure with the third planet
at 67 days. We found that the 67-day model does an overwhelmingly
better job at producing periodograms which resemble the periodogram
of the actual residuals. Our Monte Carlo results indicate a 93.6\%
probability that 67 days is the correct period.

The fourth panel of Figure 3 shows the periodogram of the residuals
from the 3-planet fit. As was found also by Mayor09, the next obvious peak to
fit is the 3.15-day one, previously reported by Mayor09. A Keplerian fit to
this peak indicates a planet in a 0.028 AU orbit with a period of 3.149 days
and minimum mass of only 1.7\,\mearth (smaller by about 10\% than that found
by Mayor09). The 4-planet fit achieves a reduced chi-squared statistic (using 14
free parameters) of
3.463 and an RMS of 2.43 \ms. The estimated FAP of the peak is
$1.9\times10^{-8}$. So, the 3.15-d planet-e announced by Mayor09 also seems
well-confirmed by the combined data set and may even be about 10\% lower in mass
than first reported.

The fifth panel down in Figure 3 shows the periodogram of the residuals to
our best 4-planet fit.  Here, there are two (nearly) equal power peaks  in the
residuals power spectrum, near 37 days and 445 days. In general, our experience
has shown that it is much harder, with a given data set, to generate coherent
power at longer periods. So, between two peaks of equal power, the one with the
longer period is usually more significant. So, we fit the 445-day peak next,
though the remaining branches of the fitting tree and final solution are not
significantly altered by fitting the 37-day peak first instead. A fit to the
445-day peak indicates a minimum-mass 6.8\,\mearth planet in a 443-day 0.770 AU
orbit. The 5-planet fit achieves a reduced chi-squared statistic (using 17 free parameters)
of 2.991 and
an RMS of 2.30 \ms. The estimated FAP of the peak is $9.5\times10^{-5}$.
This 5th planet thus appears statistically well-justified by the present data set.

The sixth panel down in Figure 3 shows the periodogram of the residuals to
the 5-planet fit. A lone dominant peak remains near 37 days. This peak shows the
extreme narrowness expected of a truly coherent signal, that, if Keplerian and
real, would have a strictly fixed period and phase for its 110 cycles spanning
the past 11 years of the data set. A fit to this peak indicates a planet of
minimum-mass 3.1\,\me, on a 36.56-day orbit of size 0.146 AU. Our best 6-planet
fit (again, assuming circular orbits) achieves a reduced chi-squared statistic
(using 20 free parameters)  of 2.506 and an RMS of 2.12 \ms.  The estimated FAP of the $\sim$37-day peak is
$2.7\times10^{-6}$. Thus, this 6th planet also seems statistically well-justified by
the present data set.

Finally, the bottom panel of Figure 3 shows the periodogram of the residuals of
the 6-planet fit. This 6-planet model leaves no remaining peaks of consequence
to fit at this time. The residual peak near 59 days has been visible all the way up
the stack of panels in Figure 3 and is apparently associated
with the yearly alias involved with the 67-day, as pointed out by Mayor09. It
has a FAP (using the definition for power in Equation 1) of only 0.186. The
phased curve at this period shows significant phase gaps in both the HARPS
and HIRES data sets due to the constraint of spectroscopic observations of bright
stars mostly receiving only bright or grey lunar time. Such phase gaps further
increase the chances of a false alarm here. A 59-day planet is also completely
dynamically untenable (even with the assumption that all orbits are circular).

We wondered how many of these planets are independently
confirmed by each data set. This is difficult to answer as the Keplerian fitting 
tree process does not hold previous planets fixed as the next planet is
optimized in the process. So we looked at running the fitting process backwards.
For each independent HARPS and HIRES data set, we subtracted our model of the
system (as listed in Table 2) from the data, giving a set of residuals. The reflex motions
corresponding to the planets in our RV model were then added back in sequentially.
The advantage to this approach is that there is no optimization and resulting parameter
drift between periodograms, and one sees the sometimes non-intuitive result of
adding a known signal. This process showed us that the characterization of the
system requires the combination of both data sets.

Figure 4 shows this reverse sequence of injecting best-fit stellar reflex motion at each Keplerian
period back into velocity residuals for each data set. The set of panels on the left
show the sequence for the HIRES data set, while the panels on the right show the same sequence
for the HARPS data set. The top panel on each side shows the periodogram of the residuals
after fitting out all 6 planets. In each successive panel, the period of the injected signal is denoted by a
red vertical tick mark.

\begin{figure}
\epsscale{0.85}
\plotone{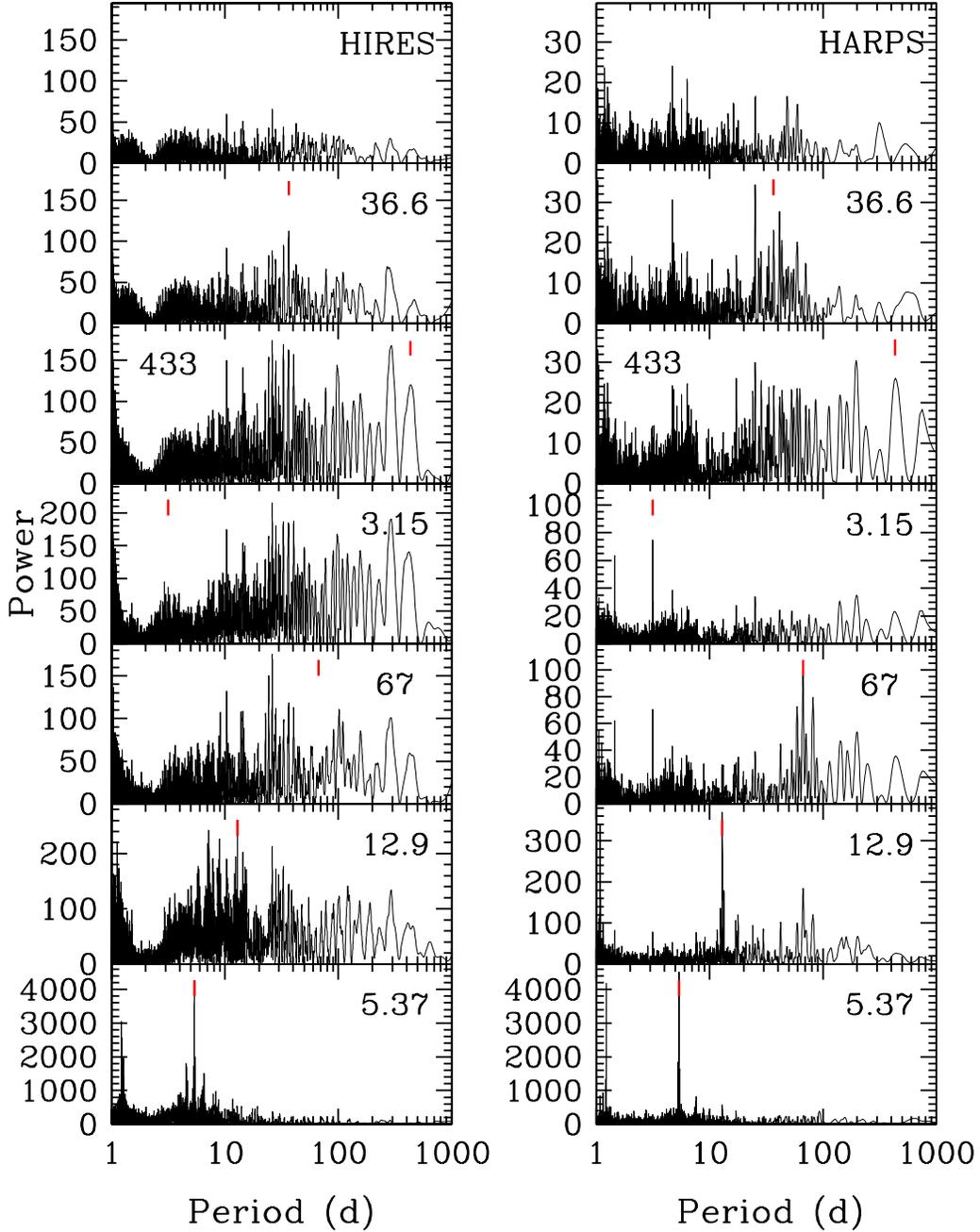}
\caption{The effect of sequentially adding signal in reverse order at each Keplerian period back into
the residuals of each data set for GJ 581. Panels on the left show the results for the HIRES
data set, while those on the right show the results for the HARPS data set. The top panels show the periodograms of the residuals from the 6-planet fit. The annotations and red vertical tick
marks in each panel indicate the period of the last injected signal prior to computing each periodogram.}
\end{figure}

The second panel on the left of Figure 4 shows the effect of injecting the 37-day signal
into the HIRES residuals. The 37-day signal is clearly visible
in the HIRES data set alone and manifests at the correct period.
The 3rd panel on the left of Figure 4 reveals that the
433-day signal is also visible and also manifests near its true period. The 4th panel on the left illustrates that
adding in the 3.15-day
signal generates power primarily at the non-intuitive period of about 26 days. The spectral window
of the HIRES sampling times has peaks at 29.53, 363.24, 1.003, and 179.72 days. This 26-day
peak could thus be drawing power from at least three sources 1) a lunar alias of the 36.6-day planet,
2) a half-year alias of the 66.9-day planet, and 3) both a one-day and a half-year alias of the 3.15-day planet.
These aliasing and sampling effects produced by the particular HIRES data time stamps 
render the 3.15-day planet inconspicuous in the power spectrum of the HIRES data taken alone.
The 5th panel on the left reveals that injecting the 67-day signal makes the situation more confusing,
by introducing more peaks. This
demonstrates that the combination of both data sets is required to see this planet
clearly, apparently because it is near an integer multiple of the lunar month which results in
difficulties getting complete phase coverage. The 6th panel on the left shows that injecting
the signal from the 12.9-day planet leads to another curious result, producing power at several other
frequencies aside from the true 12.9-day periodicity. Finally, the bottom panel on the left shows the
injection of the 5.4-day planet's signal. Here, the planet's amplitude is so large that its signal is
overwhelmingly manifested at the proper period.

For the HARPS data set alone, the 2nd panel on the right in Figure 4 shows that
injecting the 37-d signal generates power instead near 23 days when viewed through the
complex filter of time stamps and uncertainties specific to the HARPS data points. Apparently, the
HARPS data set alone is not able to reliably sense this planet. The 3rd panel on the right
illustrates that adding in the 433-day signal generates
power both near 433 and at its yearly alias near 200 days. The 4th panel on the right shows that
the injected signal from the 3.15-day planet also manifests well in the HARPS data set alone and
does not generate power at 26 days as happened with the HIRES data set. This is apparently a
result of many of their observing runs that garnered long blocks of contiguous nights with high
and sustained cadence. The 5th panel on the right shows
that the signal injected from the 67-day period shows up very well and at the expected period, flanked
also by its yearly aliases near 59 and 82 days. The 6th and 7th panels on the right show
that the signals from the 12.9-day and 5.4-day planets also manifest quite reliably in the HARPS data set alone.

So, in summary, it is clear that, although most of these planet signals do show up
independently in each data set, the situation is confused by aliasing with peaks in the spectral
window caused by the specific time stamps unique to each data set. It is really necessary to combine
both data sets to sense all these planets reliably.

A summary of our best Keplerian fit with (forced) circular orbits is presented
in Table 2. The fitted mean anomalies are
reported at epoch JD 2451409.762. The final parameters shown here are
slightly different than those quoted for the 
fits along the fitting tree and represent our best overall model. Uncertainties
(in parentheses) on each quantity are determined from
1000 bootstrap trials from which we take the standard deviations of the fitted
parameters to the bootstrapped RVs as the uncertainties. We also calculated
uncertainties with a Markov-chain Monte Carlo estimator, and both
are in good agreement. The 6-planet all-circular fit achieves a reduced
chi-squared parameter of 2.6503 and an RMS of 2.118 \ms. Allowing eccentricity
to float for any or all of the 6 planets did not produce any significant
improvement in the overall quality of the fit, either in the reduced chi-squared
statistic, in RMS, or in required stellar jitter. Given the very small amplitudes of
the signals, it is not altogether surprising that almost all of the fitted eccentricities
are statistically consistent with zero.

Our best fit indicates
that, if one allows a stellar jitter of 1.4 \ms, the reduced chi-squared statistic drops
to 1.0. This jitter estimate agrees quite well with that of Mayor09, who
found a value of 1.2 \ms\ from their 4-planet fit. Little is known about the
lower bounds of jitter for any star. If the true stellar RV jitter is even
less than this, there could yet be more planets in the system that further
precision RV data might reveal. But we also find it remarkable that this
star's jitter has not exceeded 1.4 \ms\ over the 11-year extent of the data
and that the entire data set can be fit to this level of precision by only 6
circular orbits (20 free parameters). Backing out the stellar jitter in the
quadrature sum implies that, with this data set, we are able to track the
motion of the 6 planetary companions around GJ~581 to a precision of
1.6 \ms\ over 11 years. Figure 5 shows the phased barycentric reflex
velocities of the host star due individually to each companion in the system.
Except for the 2nd panel, the ordinate scaling has been held constant to
simplify inter-comparison of the various planets.

\begin{figure}
\epsscale{0.9}
\plotone{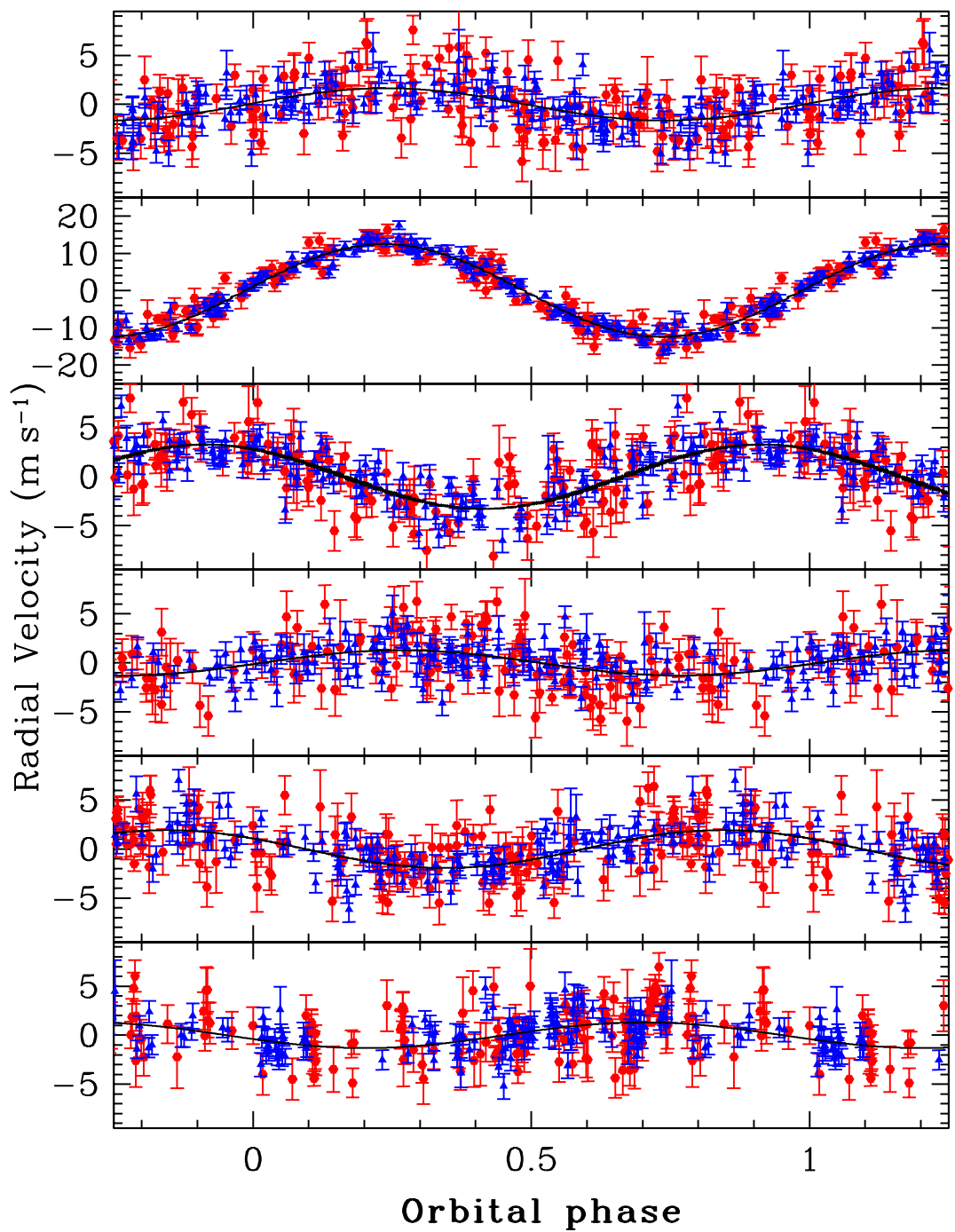}
\caption{Phased reflex barycentric velocities of the host star due
individually to the planets at 3.15 days, 5.37 days, 12.9 days, 37 days,
67 days, and 433 days from the all-circular fit of Table 2. Filled (red)
hexagon points are from Keck while filled (blue) triangles are from HARPS.}
\end{figure}

We also explored many solution sets allowing eccentricities to float for some
or all of the planets. As mentioned above, none produced any significant
improvement in overall fit quality. Moreover, most models quickly became unstable
once eccentricities rose much above 0.2 or so. Our very best eccentric fits
benefitted primarily from allowing eccentricity on the 67-day and 37-day planets' orbits
with these two planets participating in a secular resonance.

We also carefully examined the effects of including dynamics in the fitting
process. The SYSTEMIC Console includes a Gragg-Bulirsch-Stoer integrator that
can be used to model planet-planet gravitational interactions. We find that
dynamical effects have an insignificant effect on improving the fit presented
in Table 2, and the 6-planet system appears dynamically stable over at least a
50 Myr timescale. We also explored the possibility of setting limits on the
inclination of the system from dynamical stability experiments. Mayor09 had
found that the dynamical stability of their 4-planet system, particularly the
stability of the 3.15-day planet,  imposed a lower bound of about 40$^{\circ}$
for the inclination of the system (presumed co-planar). Thus, each of GJ~581's
planets could not be more massive than about 1.6 times their minimum mass.

We find that, through stability considerations, all-circular orbit solutions
only very weakly constrain the inclination of the system. Planetary masses have
to be increased by a factor $> 10$ to provoke instability in less than 50 Myr,
and that translates to a lower bound on the inclination of only
$\sim6^{\circ}$. Eccentricities do play a role in setting a lower limit to the inclination.
Floating eccentricity solutions with mass factors
($1/m\sin{i}$) $>$ 1.4 are unstable.  Even if only low eccentricities
($<0.2$) are allowed in the orbits, an upper limit for $1/m\sin{i}$ of
1.4\,--\,1.5 is indicated from dynamical stability considerations alone.
This implies that, if any of the orbits are eccentric, the system's
inclination is likely to be $> 45^{\circ}$. It seems likely that small eccentricities
are probably present in some or even all of these orbits. However, since we
cannot prove that small eccentricities are present, the inclination can't yet
really be definitively constrained.

Table~3 gives the semi-amplitudes of least-squares 
sine fits of the photometric observations (Figure 1) corresponding to each of the
radial velocity periods modeled in this paper. These upper limits to
brightness variability are all very small and supportive of Keplerian motion
of planetary companions as the cause of all the radial velocity variations.

\begin{figure}
\includegraphics[angle=-90,scale=0.8]{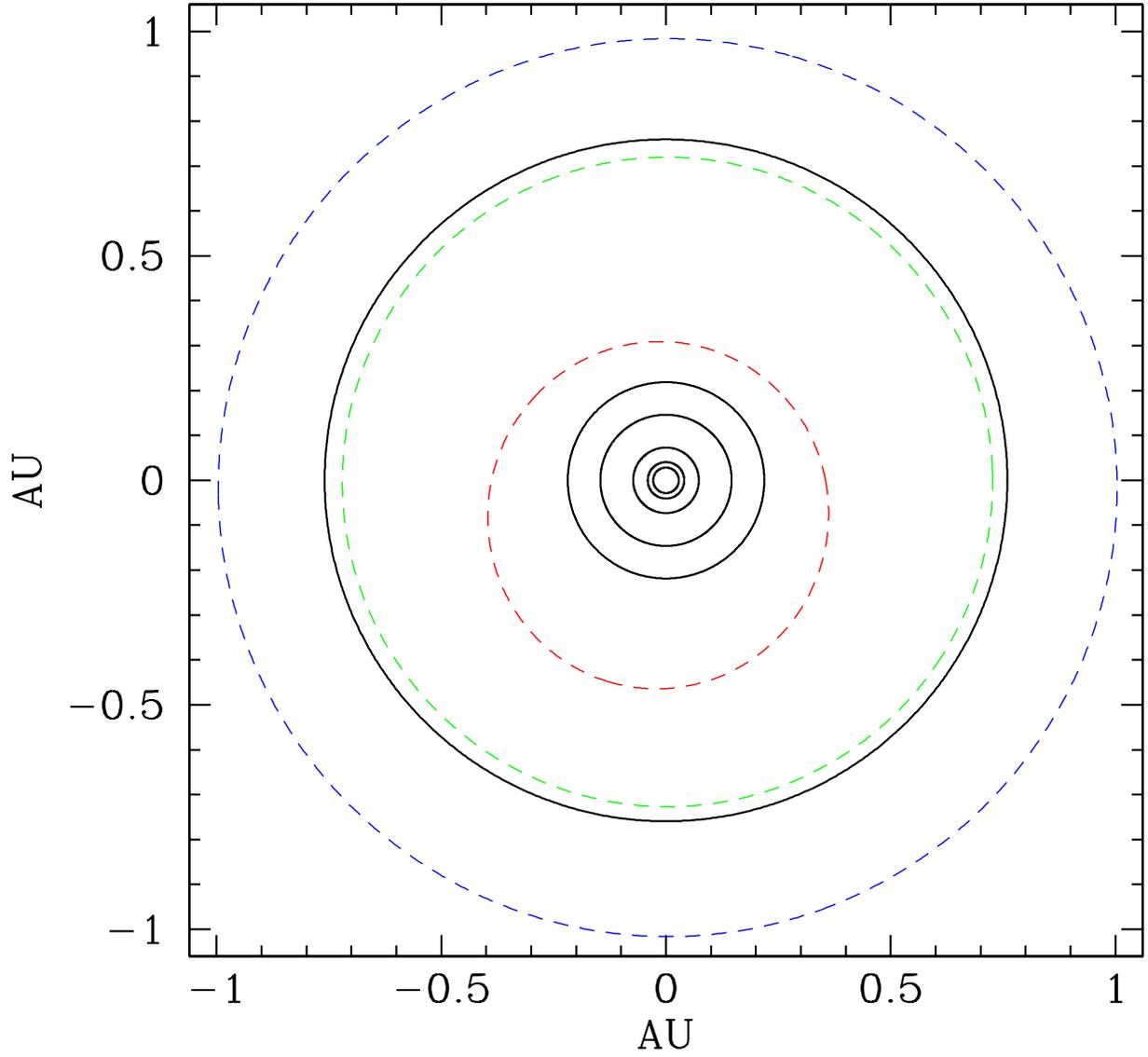}
\caption{Top view of the GJ 581 system. For reference, the orbits of Earth, Venus, and Mercury are overlaid as dashed blue, green, and red lines respectively.}
\end{figure}

Figure 6 shows a simple top view of the system, with the axes labeled in AU.
For reference, the orbits of Earth, Venus, and Mercury are overlaid as blue, green, and
red dashed lines respectively. The entire GJ 581 system would fit comfortably within the Earth's
orbit. And the basic structure of the GJ 581 system (with its nearly all-circular orbits and
a tight inner clutch of planets accompanied by a much more distant outer planet) is in
some respects eerily reminiscent of the nearly all-circular orbits of our own solar system, with
its inner clutch of terrestrial planets and attendant distant Jupiter.

\section{Characteristics of the 37-day planet}

The GJ~581 system has a somewhat checkered history of habitable planet claims,
so a brief
historical review of the alleged properties of the various planets in this
system is appropriate. Both the 12.9-day and 83-day planets reported by Udry07
were initially thought likely to be  habitable planets. However, further
analysis by others (previously described in the introduction) showed that the
12.9-day planet was likely too hot and the 83-day too cold to support
habitability. Two years later, when Mayor09 revised the period of the 83-day
planet to 67 days, that planet's prospects for habitability increased somewhat,
despite the fact that, at a minimum mass of 7.1 \me, and a maximum mass of up
to 11.4 \me, the distinction between a rocky planet and an ice-giant becomes
uncertain. The new mass, as derived here is 5.6\,--\,8.4 \me. But even with a
Bond albedo of 0, at its distance of 0.218 AU from the star, ignoring the
effects of the star's spectral energy distribution, that planet's maximum
equilibrium temperature would be only 203~K.

However, if confirmed, the 37-day planet candidate offers a
solid case for a potentially habitable planet in this very nearby system.
The best Keplerian fit to the data indicates a 3.1\,\mearth planet in a
circular 36.6-day orbit of semi-major axis 0.146 AU. The dynamical stability
investigations presented by Mayor09 also impose a lower bound on the orbital
plane inclination, constraining the upper bound on the mass of GJ~581g to be
no more than 1.6 times its minimum mass. We find a similar bound of about
1.4 assuming none of the orbital eccentricities exceed 0.2. So, the likely
mass for this planet candidate is 3.1\,--\,4.3\,\me. Using the results  of
\cite{sea07}, the radius of GJ~581g is expected to be 1.3\,--\,1.5\,\re\ if
homogeneous and composed primarily of the perovskite phase of MgSiO$_3$
(Earth-like), or 1.7\,--\,2\,\re\ if water-ice. All radii are predicted to be
$\sim$20\% smaller if the planet is differentiated, so the planet is likely to have
a radius below 1.5\,\re. The mass and radius estimates imply a surface gravity
of $\sim$ 1.1\,--\,1.7 g, very near that of the Earth.

\cite{sel07} offer a detailed summary of conditions for exoplanet habitability,
with specific reference to the GJ 581 system, but  cautioned that there are
many factors that affect habitability. Distance
from the star is but one of these factors. A planet may not have formed with
or retained sufficient water. Gravity may be too weak to hold an atmosphere
against photodissociative-escape processes. The planet might maintain an
active geological cycle to replenish atmospheric CO$_2$. Or a planet may
have accreted a massive H$_2$ - He envelope that would keep the surface
pressure too high to prevent water from existing near the surface in
liquid form. \cite{sel07} argue that avoiding the last two scenarios
requires a planet's mass to be roughly in the range of 0.5\,--\,10 \me.
GJ~581g easily satisfies this mass condition.

\cite{sel07} also make the point that a necessary and sufficient condition
for habitability is that T$_{eq}$ must be lower than about 270~K. The
equilibrium temperature \citep{sel07} is given by
T$_{eq}^4$ = L (1-A) /(16 $\pi$ a$^2$ $\sigma$), where $\sigma$ is the
Stefan-Boltzmann constant, a is the orbital radius, and A is the Bond
albedo (the fraction of power at all wavelengths scattered back into
space). This formula assumes a spherical planet with the energy that
is absorbed over the starlit hemisphere being uniformly reradiated
over the entire surface of the planet. The Bond albedo does not
however depend solely on the geometric and physical characteristics
of the planet, but also on the spectral energy distribution of the
host star. M stars emit a large amount of their radiation in the
infrared. As a result, since the greenhouse effect works by absorbing
infrared radiation, the surface temperatures would be higher than
predicted by such simple calculations. The thickness, density, and
composition of the atmosphere also significantly influence the
greenhouse effect. These in turn are ultimately influenced by the
planet's mass and radius (its surface gravity) and internal structure.
The chaotic processes that operated during the planet's formation and
its subsequent evolution determine the planet's mass, radius, and
internal structure. So the problem is complex and clearly
over-simplified by this formula.

Nevertheless, we estimate the equilibrium temperature given
L$_{\star}$ = 0.0135\,\lsun\ for the host star. We assume a Bond albedo for
the planet of A=0.3, a typical value for objects in the inner Solar System
(Earth's Bond albedo is 0.29). For the 36.6-day planet candidate, its
semi-major axis of 0.146 AU leads to an equilibrium temperature of 228~K.
If instead the Bond albedo is assumed to be 0.5, the equilibrium temperature
becomes 209~K. This planet candidate would thus appear to also satisfy
another necessary condition for habitability, that T$_{eq}$ $<$ 270~K.

An equally important consideration is the actual surface temperature T$_s$.
The equilibrium temperature of the Earth is 255~K, well-below the freezing
point of water, but because of its atmosphere, the greenhouse effect warms
the surface to a globally-averaged mean value of T$_{s}$ = 288~K. If, for
simplicity, we assume a greenhouse effect for GJ~581g that is as effective
as that on Earth, the surface temperatures should be a factor 288/255 times
higher than the equilibrium temperature. With this assumption, in the absence
of tidal heating sources, the average surface temperatures on GJ~581g would
be 236\,--\,258~K. Alternatively, if we assume that an Earth-like greenhouse
effect would simply raise the equilibrium temperature by 33~K, similar to
Earth's greenhouse, the surface temperature would still be about the same,
242\,--\,261~K. Since it is more massive than Earth, any putative atmosphere
would likely be both denser and more massive. It would be denser because of
the larger surface gravity, which would tend to hold more of the atmosphere
closer to the surface. And the atmosphere may be significantly more massive
if we simply assume that the planet went through a formation process similar
to that of the Earth and that all the bodies that went into forming GJ~581g
had the same relative amount of gasses as in the bodies that went into making
up the Earth. Some of these gases would subsequently be outgassed to make the
atmosphere. Note however, that the amount of outgassing can depend critically
on the (evolving) internal structure of the planet. More simply, the rocks
that hold the gases in GJ~581g will have experienced different pressures and
temperatures than those in the Earth. In turn, this determines how easily the
gases would be released.

Gliese 581g is likely to have evolved to a spin-synchronous configuration,
leading to one hemisphere of the planet lying in perpetual darkness.  \cite{jos97}
presented three-dimensional simulations of the atmospheres of synchronously
rotating planets in the habitable zones of M dwarfs and concluded that such
tidally-locked planets can support atmospheres over a wide range of
conditions, and despite constraints involving stellar activity, are very likely to
remain viable candidates for habitability. \cite{jos03} presented a more sophisticated
three-dimensional global atmospheric circulation model that expanded on the previous
work of \cite{jos97} and evaluated the climate of a spin-synchronous
planet orbiting an M dwarf star. The results of that study reinforced the conclusions of
\cite{jos97} that synchronously rotating planets within the circumstellar habitable zones
of M dwarf stars should be habitable.

\section{Implications for $\eta_{\oplus}$}

In recent years, the parameter $\eta_{\oplus}$ has been minted by the NASA
community to aid in evaluating and planning for space missions that seek to
discover habitable planets. The official definition of $\eta_{\oplus}$ is
given by the Exoplanet Task Force Report \citep{lun08} as:  ``The fraction of
stars that have at least one potentially habitable planet. The Task Force
defines a potentially habitable planet as one that is close to the size of the
Earth and that orbits within the stellar habitable zone. Close to Earth-sized
means between 1/2 and twice the radius of the Earth or in terms of mass between
0.1\,--\,10 times the mass of the Earth. These two definitions are equivalent
if a fixed density equal to that of the Earth is adopted.''

If confirmed, the discovery of GJ~581g, a planet of 1.3\,--\,2\,\re\ orbiting
in the habitable zone of such a nearby star implies an interesting lower limit
on $\eta_{\oplus}$ as there are only $\sim 116$ known solar-type or later stars
\citep{tur03} out to the 6.3 parsec distance of GJ~581. The definition of
$\eta_{\oplus}$ does not exclude our own Solar system from
consideration, so among that volume-limited sample out to 6.3 pc, we would
now know of two habitable systems, GJ~581 and our own
solar system, implying $\eta_{\oplus}$ is at least 2/116 or 1.7\%. But not all of
these nearest 116 stars have been under survey long enough and with enough
cadence to discern such rocky planets. The first planet found around GJ~581,
a 16.6 \mearth ice-giant,  required 20 observations to detect (Bonfils05).
The next two planets, a 12.9-day 5\mearth planet, and an 83-day 8\mearth
planet, required 50 observations over a time span of 1050 days (Udry07).
Even so, the orbital periods and minimum masses of both planets required
significant revision when additional observations by Mayor09 brought the
total to 119 over a time span of 1570 days. The two new planets presented
here required over 240 observations to discern. So it would seem that at
least $\sim$200 observations are required to reliably detect and characterize
a few-earth-mass planet in the habitable zone of a nearby K or M dwarf.

To the best of our knowledge, only $\sim$61 of these 116 nearest stars have
published evidence of being monitored by our LCES programs and/or by various
similar programs involving CPS, HARPS, CORALIE, HET, UVES, CFHT, etc., and
only 9 of these are known to us as having enough observations ($>200$) to
have a reasonable chance at being able to detect such small amplitude signals.
So, the current extent of the various RV-based exoplanet surveys implies an
incompleteness factor of 116/9 or a factor of 13 increase in the 1.7\% lower
limit, making $\eta_{\oplus}$ at least 22\%. Looking a little further out, to
10 pc, there are about 302 F, G, K, and M dwarfs. Of these, we could find
evidence in the literature for only $\sim$125 that are under survey and only
about 10 of these targeted stars that have more than 200 observations. So,
having the Sun and Gliese 581 be the only known habitable exoplanet systems
in a volume-limited sample out to 10 pc would imply a lower limit for
$\eta_{\oplus}$ of 2/302 times a survey incompleteness factor or 302/10, or
about 20\%. Looking further out still, to 12 pc, there are about 530 stars
and only about 179 under precision RV survey, with only 13 of these stars
having at least 200 observations. Those numbers translate to a lower limit
for $\eta_{\oplus}$ of 2/530 times a survey incompleteness factor or 530/13,
or about 15\%. Conclusions drawn from ever larger local volume-limited
samples have diminishing credibility as the survey incompleteness rises
dramatically with increasing stellar count with survey volume.

Another unavoidable incompleteness factor involves the random inclinations of
exoplanet orbits. Assuming random inclinations, $(1-\cos{30^{\circ}})$ or about
13\% of the stars in any volume-limited sample would be expected to have
orbital inclinations $\leq30^{\circ}$ (with respect to the plane of the sky).
Were such systems to harbor planets, their observed $K$ values would be at
least a factor of 2 less than if edge-on. For example, the $K$ value for
GJ~581g is only 1.3\,\ms. An additional factor of 2 decline in $K$ for those
13\% of similar stars that are at low inclinations (and also harbor habitable
planets) would bring the observable reflex velocity amplitude down to
0.65 \ms, at or below the expected stellar jitter for the even the quietest
stars. With today's largest telescopes and cutting-edge RV precision (1 \ms),
for stars as faint as typical nearby M dwarfs, photon statistics dominate the
error budget and, in combination with stellar jitter, make routine and
wholesale detectability of such low $K$ values extremely unlikely given the
available cadence of the present surveys. We can conservatively expect
another factor of at least 13\% incompleteness correction in our
present surveys of this volume-limited sample.

So, finding a habitable exoplanet system this soon among the nearest few
hundreds of stars in the local stellar neighborhood, in spite of the present
high level of survey incompleteness and including our own solar system also
as a habitable system implies that $\eta_{\oplus}$ could be on the order of a
few tens of percent.

\section{Summary}

We have presented 11 years of precision HIRES RV data for GJ~581.
Our 122 velocities, when combined with the 119 high-quality HARPS velocities of
Mayor09 indicate 6 companions in Keplerian motion around this star. The data strongly
confirm the 5.37-day planet-b, the 12.9-day planet-c, the 67-day planet-d, and
the 3.15-day planet-e candidates previously announced by Bonfils05, Udry07, and
Mayor09. The data also indicate two more planets in this system a 7.0\,\mearth
433-day planet and a 3.1\,\mearth 36.6-day planet. The latter orbits squarely
in the habitable zone of the star.

The National Academy of Science's recently released 2010 Astronomy and
Astrophysics Decadal report
lists "seeking nearby habitable planets" as one of its top three objectives for the
coming decade. For the past decade, the Doppler
velocity method has been the most productive
channel for planet detection. In coming years, RV detection will almost
certainly continue to delineate the closest and astrobiologically most
compelling planets, limited mostly by available telescope time. As the RV
amplitudes of truly habitable planets are
near the detection limit, collaboration between leading teams would be
extremely helpful. The planet candidate GJ~581g presented here, if confirmed, offers a
compelling case for a potentially habitable planet, but its RV signature
required the combined power of extensive HARPS + HIRES data sets.
RV precisions approaching 1 \ms, and cadences of hundreds of observations
on the quietest stars are necessary to securely detect such low-mass planets.
GJ~581 does seem to be one of those very quiet stars, with an apparent stellar
jitter of no more than 1.4 \ms. Remarkably, the star has  maintained this low
level of jitter for 11 years now.

A straightforward and very cost-effective way to realize the 2010 Decadal report's
goal of seeking nearby habitable planets, without the need to develop a
new generation of "advanced" precision
optical or infrared spectrometers, is to build dedicated 6-8 meter class
Automated Planet Finder telescopes, one in each hemisphere. Such dedicated
telescopes, instrumented with today's state-of-the-art precision
radial velocity spectrometers, like HARPS or HIRES or Magellan's new PFS
(Planet Finder Spectrometer) could, within a few short years, provide the
necessary cadences of hundreds of observations on all of the nearby quiet
G, K, and M dwarf stars within 10 pc, in all probability revealing many other
nearby potentially habitable planets. Riding on the coat tails of existing engineering by
closely copying the Magellan 6.5-m telescopes, each facility could probably be built
(and instrumented with a precision RV spectrometer) for about \$50 million, or \$100 million
total for telescopes in both hemispheres. Indeed,
if $\eta_{\oplus}$ is really as high as several tens of percent (or is even only no more
than a few percent) having only a single planet finder in one hemisphere could accomplish
pretty much the same goal, for a mere \$50 million. With this single capital investment,
one could make sure, swift, and cost-effective progress on one of the 2010 Decadal
report's three primary science goals.

Finally, it is important to keep in mind that, though all 6 planets presented here
are well-supported by the calculated reduced chi-squared
statistics and also by several different variants of FAP statistics, and the entire
6-planet system is consistent with the combined data set from both teams,
caution is warranted as most of the signals are small. And there may yet be unknown
systematic errors in either or both data sets. For example, \cite{pon10} have
recently concluded from a detailed analysis of HARPS CoRoT-7 data that "On
the whole, there is a mounting body of evidence that unexplained variations at
the 5-10 \ms\ level may exist in HARPS RVs for targets in the brightness range
of CoRoT-7." GJ 581 is only about a magnitude brighter that CoRoT-7, so it may not
be completely out of the question that HARPS data for GJ 581 might also be affected by
such unexplained errors. And to be completely fair, the HIRES data set could also have
undiscovered systematic errors lurking within. This is very difficult work and there is
no shame or dishonor in uncovering residual systematic errors at these levels of precision.
Collegial and unabashed inter-team comparisons
on stars like GJ 581 and GJ 876 will be crucial to quantifying the true precision limits
of any team's data sets. Finally, because of the very small amplitudes involved, allowing
significant eccentricities into the Keplerian fitting tree may yield viable alternate solutions.
Here, phase gaps in data sets become problematical as fitting routines generally
allow eccentricity to utilize these gaps, driving up the eccentricity artificially to
enhance the quality of the fit, and hiding much of the velocity swing from eccentricity in the
phase gap. Such situations sometimes result in misleading solutions that
can overlook or mask additional planets in the system.

Confirmation by other teams through additional high-precision RVs would be
most welcome. But if GJ~581g is confirmed by further RV scrutiny, the mere
fact that a habitable planet has been detected this soon, around such a
nearby star, suggests that $\eta_{\oplus}$ could well be on the order of a few tens
of percent, and thus that either we have just been incredibly lucky in this early
detection, or we are truly on the threshold of a second Age of Discovery.

\acknowledgments

SSV gratefully acknowledges support from NSF grant AST-0307493. RPB gratefully
acknowledges support from NASA OSS Grant NNX07AR40G, the NASA Keck
PI program, and from the Carnegie Institution of Washington. NH acknowledges
support from the NASA Astrobiology Institute under Cooperative Agreement 
NNA04CC08A at the Institute for Astronomy, University of Hawaii, and NASA 
EXOB grant NNX09AN05G. GWH and MHW acknowledge support by NASA,
NSF, Tennessee State University, and the State of Tennessee through its
Centers of Excellence program. The work herein is based on observations obtained
at the W. M. Keck Observatory, which is operated jointly by the University of California
and the California Institute of Technology, and we thank the UC-Keck and NASA-Keck
Time Assignment Committees for their support. We also acknowledge the contributions
of fellow members of our previous California-Carnegie Exoplanet team in helping to obtain
some of the earlier RVs presented in this paper. We also wish to extend our special thanks
to those of Hawaiian ancestry on whose sacred mountain of Mauna Kea we are privileged
to be guests. Without their generous hospitality, the Keck observations presented herein
would not have been possible. Finally, SSV would like to extend a very special thanks to
his wife Zarmina Dastagir for her patience, encouragement, and wise counsel. And even
though, if confirmed, the habitable planet presented herein will officially be referred to by
the name GJ 581g, it shall always be known to SSV as "Zarmina's World".

{\it Facilities:} \facility{Keck}.

\clearpage
\begin{deluxetable}{rrrr}
\tablenum{1}
\tablecaption{Radial Velocities for GJ~581}
\label{vel43848??}
\tablewidth{0pt}
\tablehead{
JD & RV & error \\
(-2450000)   &   (\ms) & (\ms)
}
\startdata
1409.76222 & 6.96 & 1.89\\
1586.14605 & -10.24 & 3.22\\
1704.91213 &  0.47 & 2.89\\
2003.95507 & -4.37 & 3.65\\
2100.86678 & -19.45 & 2.22\\
\enddata
\tablecomments{Table 1 is presented in its entirety in the electronic edition
of the Astrophysical Journal.  A portion is shown here for guidance regarding
its form and content.}
\end{deluxetable}

\begin{deluxetable}{rlllllllll}
\tabletypesize{\scriptsize}
\tablenum{2}
\tablecaption{Orbital Parameters for GJ~581 Planet Candidates}
\label{candid}
\tablewidth{0pt}
\tablehead{
\colhead{Planet}  & \colhead{Period} & \colhead{$K$} & \colhead{$m\sin{i}$} & \colhead{$a$} & \colhead{Mean Anomaly\tablenotemark{a}}
\\
\colhead{} & \colhead{(days)} & \colhead{(\ms)} & \colhead{(\me)}  & \colhead{(AU)} & \colhead{($^{\circ}$)} & \colhead{FAPS} }
\startdata
b & 5.36841  (0.00026) & 12.45 (0.21) & 15.6 (0.3) & 0.0406163 (1.3e-6) & 276.1 (4.9) & 6.8e-306 \\
c &  12.9191  (0.0058) & 3.30  (0.19) & 5.6 (0.3)  & 0.072993 (2.2e-5)  & 33 (19)     & 2.3e-33 \\
d &  66.87 (0.13)      & 1.91 (0.22)  & 5.6 (0.6)  & 0.21847 (2.8e-4)   & 56 (27)     & 2.5e-6 \\
e &  3.14867 (0.00039) & 1.66 (0.19)  & 1.7 (0.2)  & 0.0284533 (2.3e-6) & 267 (40)    & 1.9e-8 \\
f &  433 (13)          & 1.30 (0.22)  & 7.0 (1.2)  & 0.758 (0.015)      & 118 (68)    & 9.5e-5 \\
g &  36.562 (0.052)    & 1.29 (0.19)  & 3.1 (0.4)  & 0.14601 (1.4e-4)   & 271 (48)    & 2.7e-6 \\

\enddata
\tablenotetext{a}{The fitted mean anomalies are reported at reference epoch JD 2451409.762.}
\end{deluxetable}

\begin{deluxetable}{ccc}
\tablenum{3}
\tablewidth{0pt}
\tablecaption{Photometric Semiamplitudes Modulo the Radial Velocity Periods}
\tablehead{
\colhead{} & \colhead{Planetary Period} & \colhead{Semi-amplitude} \\
\colhead{Planet} & \colhead{(days)} & \colhead{(mag)}
}
\startdata
 b & 5.36841 & $0.00045\pm0.00044$ \\
 c & 12.9191 & $0.00083\pm0.00044$ \\
 d & 66.87   & $0.00129\pm0.00044$ \\
 e & 3.14867 & $0.00061\pm0.00045$ \\
 f & 433     &       \nodata       \\
 g & 36.562  & $0.00058\pm0.00047$ \\
\enddata
\tablecomments{The data set is insufficient to address the 433 day period.}
\end{deluxetable}

\end{document}